\documentclass[twocolumn,showpacs,preprintnumbers,superscriptaddress,amsmath,amssymb,prl]{revtex4-1}

\usepackage{graphicx}
\usepackage{dcolumn}
\usepackage{bm}
\usepackage{amssymb}
\usepackage{multirow}

\newcommand{\1}{\begin{equation}}
\newcommand{\2}{\end{equation}}
\newcommand{\ea}{\begin{eqnarray}} 
\newcommand{\ee}{\end{eqnarray}}

\newcommand{\Sum}[2]{{\sum\limits_{#1}^{#2}}}

\begin{document}

\title{Interaction Induced Directed Transport in AC-Driven Periodic Potentials}

\date{\today}


\author{Benno Liebchen}
\email[]{Benno.Liebchen@ed.ac.uk}
\affiliation{SUPA, School of Physics and Astronomy, University of Edinburgh, Edinburgh EH9 3FD, United Kingdom}
\author{Peter Schmelcher}
\email[]{Peter.Schmelcher@physnet.uni-hamburg.de}
\affiliation{Zentrum f\"ur Optische Quantentechnologien, Universit\"at Hamburg, Luruper Chaussee 149, 22761 Hamburg, Germany}%
\affiliation{The Hamburg Centre for Ultrafast Imaging, Universit\"at Hamburg, Luruper Chaussee 149, 22761 Hamburg, Germany}

\begin{abstract}
We demonstrate that repulsive power law interactions can induce directed transport of particles in dissipative, ac-driven periodic potentials, 
in regimes where the underlying noninteracting system exhibits localized oscillations. 
Contrasting the well-established single particle ratchet mechanism, 
this interaction induced transport is based on the collective behaviour of the interacting particles yielding a 
spatiotemporal nonequilibrium pattern comprising persistent travelling excitations. 
\end{abstract}

\maketitle

\begin{figure}[htb]
\begin{center} 
\includegraphics[width=0.49\textwidth]{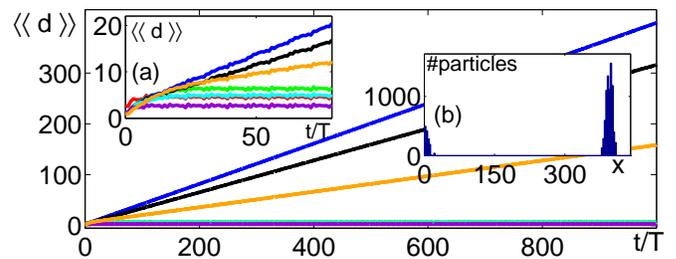}
\caption{Time evolution $\langle\langle d \rangle\rangle (t)$ for $N=6$ (ochre: $N=20$), averaged over $10^4$ ensembles, for
$\alpha=0$ (red, 2nd lowest); $\alpha=1$ (green, 4th lowest); $\alpha=2$ (blue, highest); $\alpha=5$ (black, 2nd highest; ochre 3rd highest); $\alpha=20$ (cyan, 3rd lowest); $\alpha=175$ (purple, lowest). 
Inset (a): Magnification of the short time behaviour of $\langle\langle d(t) \rangle\rangle$. Inset (b): Distribution of 
$\langle d \rangle$ for $\alpha=5.0$ and 10000 ensembles
at $t=1000 (2\pi/\omega_0)$. Parameters: $a=m=k=1.0; V=25; \gamma=8.6; v_0\in(-250,250)$; $\omega_0=3.53605$ 
($2\omega_0=\sqrt{2Vk^2/m}$).
}
\label{fig1}
\end{center}
\end{figure}

\paragraph*{Introduction}
The ratchet effect describes the emergence of a directed particle transport due to an interplay of symmetry breaking 
\cite{prost94,flach00,denisov14} and nonlinearity
in a system where all applied forces and gradients vanish after averaging over time and space 
\cite{magnasco93,flach00,schanz01,reimann02,hanggi09,denisov14}. 
Originally inspired by the desire of understanding how molecular motors convert chemical 
energy into directional motion \cite{magnasco93}, 
ratchets have now advanced to a widespread and topical paradigm with applications
in various branches of nonequilibrium atomic \cite{mennerat99,schiavoni03,salger09,renzoni09,denisov14}, 
condensed matter \cite{faucheux95,marquet02,matthias03,hanggi09,roeling10,hanggi11,drexler13} 
and biophysics \cite{duke98,chou99,huang03,mahmud09,dileonardo10,franken12}.
Specifically, modern atomic cooling and trapping techniques \cite{pethick08,bloch08} allow for particularly clean and versatile 
realizations of ratchets in optical lattices \cite{mennerat99,schiavoni03,gommers05,gommers06,renzoni09,salger09}.
Since the established ratchet effect  
is of single particle character that can survive interactions, 
most works focused on the non-interacting single particle regime (for reviews see \cite{reimann02,hanggi09,renzoni09,denisov14})
or on continuous models with infinitely many degrees of freedom and effectively linear or Kuramoto-type interactions
\cite{reimann02,braun04,cuevas14}.
However, many relevant setups, such as driven 
dipolar atoms or molecules \cite{lahaye09,lewenstein12}, charged colloids \cite{bohlein12a} or microscopic ions \cite{pruttivarasin11} 
experience power-law (Coulomb and dipole-dipole) interactions, evoking the question of their impact on directed currents. 
For the dilute Hamiltonian regime, it was already shown that these interactions can induce intriguing dynamical reversals of the 
transport direction \cite{liebchen12}. 
\\In this work, we demonstrate that repulsive power law interactions can even
be responsible for the emergence of a directed particle transport.
Considering ions or dipoles in dissipative and laterally oscillating lattice potentials, 
interaction induced currents appear in parameter regimes where frictional energy losses suppress any 
directed current in the corresponding noninteracting system. 
The physical origin of these currents
is an interaction induced collective behaviour of the particles which 
is represented by a 
spatiotemporal nonequilibrium pattern comprising persistent travelling excitations and
has to be carefully distinguished from the well established
single particle ratchet mechanism.
\paragraph*{Setup}
We consider $N$ repulsively power law interacting point particles in one dimension
with coordinates $x_j$ and mass $m$ that are exposed to a frictional force $F_{\rm R}=-\gamma \dot x$ and a spatially periodic and laterally oscillating lattice potential 
$V(x,t)=V_0 \cos^2[kx-f(t)]$. Here $L=\pi/k$ is the distance between two adjacent minima and 
$f(t)=a \left[\cos(\omega_0 t) + \sin(2\omega_0 t) \right]$ is the biharmonic driving law, with
frequency $\omega_0$ and amplitude $a$. 
The dynamics of this model is described by $N$ coupled nonlinear, non autonomous Newtonian equations of motion:
\ea
m \ddot x_j &+& \gamma \dot x_j + V_0 k \sin\left\{2 k \left[x_j - f(t)\right]\right\} - 
\Sum{k=1, k\neq j}{N} F_{jk} = 0 \nonumber \\
F_{jk}&=&\alpha\frac{x_j-x_k}{|x_j-x_k|^{r+2}} \label{eq1}
\ee
This model might describes charged ($r=1$) or dipolar particles ($r=2$) in an ac-driven (optical) lattice potential
including a dissipative frictional force.
\\To study transport phenomena at a fixed particle density we employ periodic boundary conditions $x+L N\cong x$. We incorporate the 
corresponding Ewald sums for the interaction forces $F_{jk}$ in our simulations \cite{frenkel01},
but note that all results presented below remain qualitatively valid when truncating the interaction beyond next
neighbour ones.
\\Physically, we can distinguish four different forces determining the dynamics of a given particle ensemble:
$(i)$ the force exerted by the instantaneous static lattice ($F_{\rm s}:= - V_0k\sin(2kx) \sim V_0 k$).
If dominant, this force drags the particles along the minima of the oscillating lattice.
$(ii)$ The time dependent pseudo force which acts 
on a particle in the comoving coordinate system of the lattice ($F_{\rm o}:=-m\ddot f(t)-\gamma \dot f(t) \sim (-a \omega_0) \sqrt{\gamma^2 + (m \omega_0)^2}$). 
$(iii)$ The frictional forces $F_{\rm f}=-\gamma \dot x \sim 
\gamma \omega_0 L$ slowing down the particles in favour to keep them at fixed positions in the laboratory coordinate system.
$(iv)$ The repulsive interaction forces $F_{jk} \sim \alpha/L^{r+1}$. 
In the following parameters are chosen such that $F_{\rm s}, F_{\rm o}, F_{\rm f}$ and in many cases also $F_{jk}$ are comparable 
for sufficiently long times, i.e. none of these forces can be treated as a small perturbation.
To be in a regime where the driving forces and the
nonlinear components of $F_{\rm s}$ as well as the power law shape of $F_{jk}$ play a significant role, 
$\omega_0$ is chosen close to the resonance of the underlying linearised single particle
problem (a damped driven harmonic oscillator), i.e. $2\omega_0 \sim \sqrt{2Vk^2/m}$.
This way, we are
far from the comparatively simple regimes of adiabatically slow and fast driving resulting in effectively time independent
problems.
\\In the following we investigate the impact of different interaction strengths $\alpha>0$ on the
transport properties of spatially uniform initial states with $x_j(t=0)=L(j-N/2+1/2)+a$ ($N$ even and $j=0,1...N-1$) and random velocities $v_j(t=0)$, uniformly chosen in an interval $(-v_0,v_0)$.
To avoid capture by specific low-energy asymptotic states (attractors) we choose $v_0$ to be large, such that frictional forces dominate at short time scales. 
Experimentally, different values of $\alpha$ can of course be achieved indirectly by tuning other 
parameters such as $\omega_0,V_0$ or $\gamma$ that are easier to control.
\paragraph*{Results}
We integrate Eq.~(\ref{eq1}) using the Runge-Kutta-Dormand-Prince integration 
scheme \cite{dormand80} and calculate 
the time evolution of the ensemble averaged position $\langle d(t)\rangle:=(1/N) \sum x_j(t)$ 
(without setting $x+NL$ back to $x$).
Although noise is of course naturally present in dissipative many particle systems due to the 
fluctuation-dissipation theorem \cite{callen51,zwanzig01}, we do not include it explicitly in our simulations 
but simply note that our results are indeed robust with respect to weak (Gaussian) noise. 
Fig.~\ref{fig1} and its inset (a) show the time evolution of $\langle\langle d(t)\rangle\rangle$, where $\langle\langle . \rangle \rangle$
refers to the averaging over many $N$-particle ensembles.
For vanishing interactions ($\alpha=0$) as well as for weak interactions ($\alpha=1.0$) $\langle\langle d(t)\rangle\rangle$ 
saturates after a short initial growth at $t/T\sim 20$ and we observe that all particles asymptotically perform pure on site oscillations. 
Increasing the strength of the repulsive interactions, one might expect that $\langle\langle d(t)\rangle\rangle$ 
saturates even earlier as repulsive interactions restrict the mobility of the particles dynamics even further thereby
impeding the emergence of any single particle based transport in the system. 
However, for $\alpha=2; 5$ $\langle\langle d(t)\rangle\rangle$ grows linerly, i.e. we observe a 
permanent directed current of the underlying ions that emerges spontaneously when the repulsive interaction exceeds a certain 
threshold. 
Fig.~\ref{fig1}, inset (b) shows that apart from a few ensembles that stay close to their initial positions, 
after 1000 oscillations of the lattice $\langle d(t)\rangle$ is very similar for most ensembles
which propagate with similar velocities through the lattice.  
For stronger interactions ($\alpha=20; 175$) $\langle\langle d(t)\rangle\rangle$
saturates quickly as for $\alpha=0$, i.e. the directed current dies out after a short initial drift and each of the underlying ions 
is trapped on its lattice site. 
Hence, we observe a directed current that is interaction induced but does not survive for strong interactions. 
\begin{figure}[htb]
\begin{center} 
\includegraphics[width=0.49\textwidth]{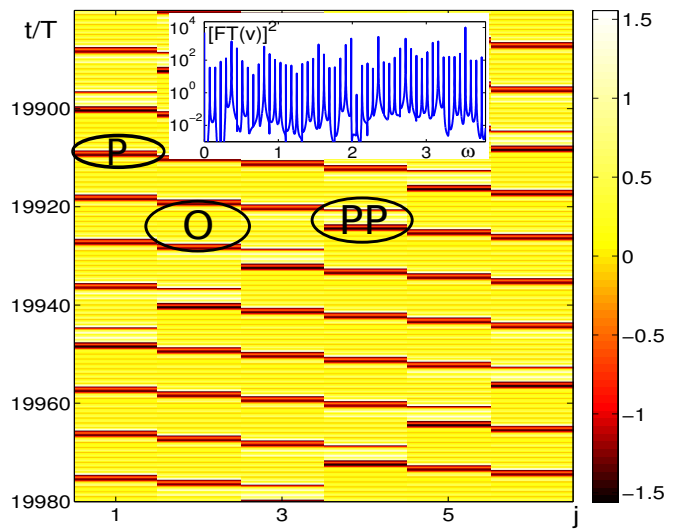}
\caption{Asymptotic trajectories $x_j(t)$ (color/grey scale denotes the value of $x_j(t)-L/2$ modulo $L=\pi$) over particle number $j=1,2...6$ and time $t/T$.
Inset: Low frequency extract of the squared Fourier spectrum of a typical $v_j(t):=\dot x_j(t)$ trajectory. 
Parameters as in Fig.~\ref{fig1} ($N=6,\alpha=5$).
}
\label{fig2}
\end{center}
\end{figure}
\\It is illustrative to further resolve this effect on a trajectory level (Fig.~\ref{fig2}). 
While for all non transporting cases ($\alpha=0; 1; 20; 175$) we asymptotically obtain exclusively on site oscillations (not shown), 
for $\alpha=5.0$ the transport is represented by a spatiotemporal pattern.
The latter consists of phases of on-site oscillations (exemplarily encircled and marked with an 'O') that are interrupted by propagation processes 
from one lattice site to its right neighbour ('P') 
and by propagation processes of two lattice
sites ('PP'). 
The spatiotemporal pattern of the on-site oscillations and propagation processes is reminescent of a situation where 
all particles oscillate approximately in phase while an additional travelling collective excitation represents the propagation process
leading to directed transport.
\\A Fourier analysis of the transporting pattern (inset in Fig.~\ref{fig2}) reveals that it is of quasi-periodic character persisting asymptotically. 
The low frequency part of the Fourier spectrum shows subharmonic frequencies of $n \omega_0/48$ with $n=1,2,3...$ 
and $2\omega_0=\sqrt{2Vk^2/m}$ reflecting the collective nature of the interaction induced directed transport. 
\begin{figure}
\begin{center} 
\includegraphics[width=0.49\textwidth]{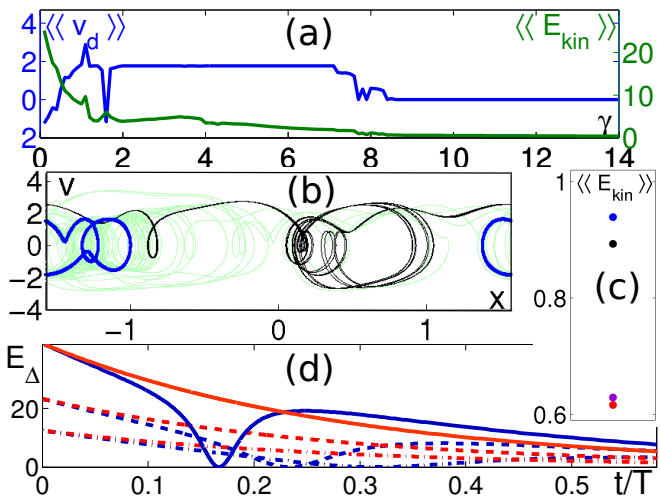}
\caption{(a): Asymptotic single particle ($\alpha=0$) transport velocity $\langle \langle v_d \rangle \rangle$ 
(blue, upper curve) and kinetic energy per particle
$\langle \langle E_{\rm kin}\rangle \rangle=(m/2)\langle \langle 
\dot x^2 \rangle \rangle$ (green, lower curve) as a function of $\gamma$, 
respectively averaged over $10^4$ single particle ensembles and the time interval $t/T\in (500,1000)$. 
Remaining parameters as in Fig.~\ref{fig1}.
(b): Projection of the asymptotic quasiperiodic dynamics of representative, exemplary trajectories in the $(x,v,t)$-single particle phase space onto the $x-v$ plane ($x$ taken modulo $L=\pi$)
for $\gamma=8.0, \alpha=0$ (black, highest saturation), $\gamma=8.6, \alpha=0$ (blue), $\gamma=8.6, \alpha=5$ (green, lowest saturation). 
(c): Asymptotic kinetic energies per particle $\langle \langle E_{\rm kin} \rangle \rangle/N$ for $N=6$ and 
$\alpha=0$ (red); $\alpha=1$ (green); $\alpha=2$ (blue); $\alpha=5$ (black); $\alpha=20$ (cyan); $\alpha=175$ (purple). 
The red dot is lying on top of the green and the cyan one. 
(d): Time evolution of the kinetic energy $E_\Delta:=(m/2)\dot \Delta^2 (t)$ 
for $\alpha=0$ (red, curves without dip), $\alpha=5.0$ (blue),
$\Delta(t=0)=2.25$ and different $\dot \Delta(t=0)<0$ (line styles). $m=\gamma=1.0$.}
\label{fig3}
\end{center}
\end{figure}
\paragraph*{Analysis}
We now focus on the mechanism underlying the interaction induced directed currents. 
Since our nonlinear system is driven out of equilibrium and breaks both parity- and time-reversal symmetry it fulfills 
the general criteria for observing a directed current \cite{denisov14}.
Indeed, in the single particle regime ($\alpha=0$ or $N=1$) a directed current can be observed, 
as long as $\gamma$ is sufficiently small, i.e. for weak dissipation (Fig.~\ref{fig3}a; blue curve).  
This current is represented by one or several coexisting
transporting limit cycle attractors (Fig.~\ref{fig3}b; black) in the underlying single particle phase space ($x,v,t$).
Physically such attractors exist as 
long as the energy transfer from the oscillating lattice to the particles can periodically 
balance the frictional energy loss, which often happens within a single period of the driving.
The slowest possible directed motion which allows for such a balancing is typically a motion of a distance $L$ 
for a driving period $T$.  In synchronization theory this is a 1:1 (or a $k:k$, $k \in \mathbb{N}$) phase locking
resonance between the particle motion and the lattice oscillation \cite{salerno00,pikovsky01}. 
A slower motion of e.g. only $L/2$ within $T$ would correspond to a particle motion which alternatingly gains and looses energy to the lattice and 
does not allow for a balancing of the energy. 
Note that the existence of such a minimal transport velocity
is in strong contrast to both the Hamiltonian regime ($\gamma=0$) and the regime of strong noise (high temperature/low particle mass) where the 
directed transport emerges as an asymmetric (chaotic deterministic or Brownian) diffusion which can generally become arbitrarily slow \cite{reimann02}. 
In the considered dissipative system, the minimal asymptotic, time-averaged transport velocity of one particle in the lattice 
that is typically possible is $v_d=\omega_0/(2k)\approx 1.77$, 
corresponding to the long plateau in (Fig.~\ref{fig3}a; blue) between $\gamma \approx 2.1$ and $\gamma\approx 7.3$.
When $\gamma$ becomes sufficiently large, the transporting attractors (Fig.~\ref{fig3}b; black) 
disappear and we have exclusively non transporting attractors (Fig.~\ref{fig3}b; blue) in the underlying phase 
space.
Values for $\langle \langle v_d \rangle \rangle$ that are smaller than $\omega_0/(2k)$ in Fig.~\ref{fig3}a are facilitated by 
non transporting attractors that coexist with transporting
ones. Some initial conditions then lead to on-site oscillations while others lead to transport with $\langle v_d \rangle=\omega_0/(2k)$.
Larger transport velocities are possible due to higher resonances such as the peak at $\gamma \approx 1.1$ representing
particles that move 4$L$ within 2$T$, i.e. a 4:2 phase locking.
\\In contrast to the single particle transport that is based on the existence of 
transporting attractors with a certain minimal velocity, 
the presence of interactions allows for the emergence of directed currents as a collective phenomenon.
Specifically, while for $\alpha=0$ the Fourier spectrum consists of peaks at $\omega_0$ and its higher harmonics $2\omega_0;3\omega_0...$ (not shown)
for $\alpha=5$ the Fourier spectrum (inset in Fig.~\ref{fig2}) shows 
subharmonic frequencies representing the possibility of a directed current of the interacting ensemble without requiring that the underlying particles 
propagate by (at least) one lattice site per driving period.
Indeed, in Fig.~\ref{fig1} the transporting ensembles propagate with $\langle v_d \rangle = L/(8T)=\omega_0/(16k)<L/T=\omega_0/(2k)$.
On the level of the $2N+1$ dimensional $N$ particle phase space, the 
corresponding emerging collective current (Fig.~\ref{fig1})
is represented by a transporting attractor (Fig.~\ref{fig3}b; green). 
For visualizing this high dimensional attractor we exploit the fact that all 
$N$ particle trajectories underlying the corresponding transporting pattern (Fig.~\ref{fig2})
are identical except for certain phase shifts, i.e. the single trajectory shown in Fig.~\ref{fig3}b is representative 
for the complete $N$ particle attractor. 
\\Let us further illuminate the microscopic mechanisms underlying the interaction induced transport.
The interaction among the particles can, in the course of the dynamics, reduce the loss of kinetic energy due to friction, which in turn
of course facilitates the emergence of directed currents.
Consider the time evolution of the kinetic energy $E_\Delta(t):=m\dot \Delta^2/2$ corresponding to the 
relative motion $\Delta:=x_2-x_1$ of a free 
frictious two-particle system $m\ddot x_{1,2} + \gamma \dot x_{1,2} \pm \alpha/[x_1-x_2]^2=0$.
While the frictional energy loss concerning the center of mass dynamics $E_x(t):=m\dot x^2/2$ with $x:=(x_1+x_2)/2$
is of course independent of the value of $\alpha$, $E_\Delta(t)$ initially decreases faster for $\alpha>0$
as compared to $\alpha=0$ (see Fig.~\ref{fig3}d) but then grows again and becomes larger than for $\alpha=0$ (where $E_\Delta(t) \propto \exp{[-2\gamma t/m]}$). 
The zero in $E_\Delta (t)$ (see blue curves in Fig.~\ref{fig3}d) hereby indicates the close encounter ie. collision of two particles.
Accordingly, a particle travelling from one lattice site to an adjacent one, can reach the
latter with a higher kinetic energy when encountering a collision on its path. 
Physically this collision induced reduction of frictional energy losses is based on the fact that upon a 
collision kinetic energy is converted into interaction energy which is not subject to frictional losses. 
In the case of the complete $N$ particle system in the oscillating lattice, 
this mechanism is still present and leads to a substantial increase of the asymptotic kinetic energies (Fig.~\ref{fig3}c) for 
$\alpha=2; 5$ (transport) as compared to the energies in the noninteracting $\alpha=0$ and non transporting $\alpha=1; 20; 175$ cases.
In these non transporting cases the particles asymptotically exhibit 
almost equidistant on site oscillations.
\\We now discuss why the interaction induced directed transport is present 
only in some regime of the interaction strength (e.g. for $\alpha=2; \alpha=5$)
but vanishes for comparatively large values of $\alpha$ ($\alpha=20; \alpha=175$).
Clearly, in the limit $\alpha \rightarrow \infty$ of a stiff ion chain where all $N$ particles exhibit an identical 
dynamics, the $N$ particle transport breaks down at the same friction strength $\gamma$ as in the single particle case ($N=1$ or $\alpha=0$).
Using linear stability analysis, we show that the $N$ particle dynamics reduces to that of a stiff ion chain already for finite values of $\alpha$, 
no matter how strong the driving is. 
Consider a two particle system (Eq.~(\ref{eq1}) for $N=2; r=1$) with periodic boundary conditions and only next neighbour interactions.
Transforming $x_i\rightarrow \tilde x_i:=2k[x_i-f(t)]$ and then to center of mass $x:=(\tilde x_1+\tilde x_2)/2$
and relative $\Delta:=\tilde x_2-\tilde x_1$ coordinates yields after 
inserting 
$\Delta = 2\pi+\delta$ ($2\pi$ is the equilibrium distance) and truncating in linear order $\delta$:
\ea
0 &=& m\ddot x + \gamma \dot x - 2V_0k^2 \sin(x) + F(t) \label{com1} \\
0 &=& m\ddot \delta + \gamma \dot \delta + 2V_0k^2 \delta \cos(2x)+8k^3\alpha \delta/\pi^3 \label{com} \\
F(t) &=& 2k \left[m \ddot f(t) + \gamma \dot f(t) \right] \nonumber
\ee
As $\cos(2x)\in(-1,1)$ it follows from the linear Eq.~(\ref{com}) that for $\alpha > V_0 \pi^3/(4k)$ the asymptotic state for 
$\delta(t)$ is uniquely determined by a fixed point attractor at $\delta=0$ with a global 
basin of attraction. Then, the state where the two considered ions exhibit an identical phase locked dynamics is stable 
(for arbitrary values of $a,\omega_0$ and $\gamma$) and any initial state
sufficiently close to it asymptotically approaches it. 
This result can be generalized to the $N$ particle case, yielding 
$\alpha > K V_0 \pi^3/(Q k)$ with $Q=4;3;2;(5-\sqrt{5})/2;1$ for $N=2,3;...6$ and $Q>0$ for $N< \infty$.
In fact, for $\alpha>K V_0 \pi^3/(Q k)$ we numerically observ that the $N$ particle ensemble discussed in the context
of Fig.~\ref{fig1} generically comes close to 
a translationally invariant state (after redistributing the nonuniform initial kinetic energies) and we observe a breakdown of the 
$N$ particle current at  the same value of $\gamma$ as in the single particle case.
The uniform (or phase locked) $N$ particle dynamics is then described by the equation of motion of its 
center of mass coordinate $x:=(1/N)\sum x_i$ (Eq.~\ref{com1}), representing a biharmonic version of an ac-driven damped physical pendulum
which is known to exhibit a period doubling route to chaos \cite{baker96}. 
Note that the criterion $\alpha >K V_0 \pi^3/(Q k)$ represents only an upper limit for the stability of the spatially uniform 
dynamics; in practice stability can of course occur already for smaller values of $\alpha$.
\paragraph*{Irregular transport}
We finally demonstrate that the interaction induced directed currents are by no means restricted to the regular regime as
represented by the propagating spatiotemporal pattern in Fig.~\ref{fig2}. 
Fig.~\ref{fig4} shows $\langle\langle d(t)\rangle \rangle$ in a regime where the dynamics of the underlying ions is irregular (Fig.~\ref{fig4}, inset a).
\begin{figure}
\label{twopart}
\includegraphics[width=0.49\textwidth]{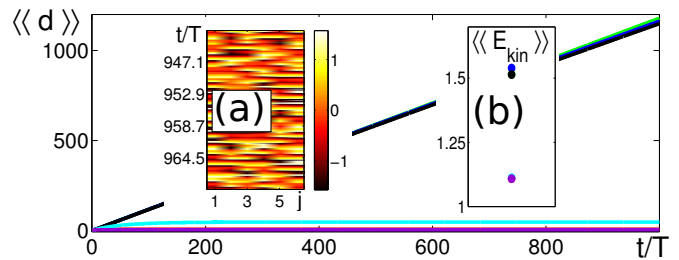}
\caption{
Time evolution $\langle\langle d \rangle\rangle (t)$ for $N=6$, averaged over $10^4$ ensembles for 
$\alpha=0.25; 2; 5$ (three upper curves in green, blue and black, lying almost on top of each other);
$\alpha=20$ (cyan, second lowest curve); $\alpha=175$ (purple, lowest curve) and $\alpha=0$ (covered by the purple curve).
Inset (a): Like Fig.~\ref{fig2}.
Inset (b): Like Fig.~\ref{fig3}c.)
Upper dots: $\alpha=0.25$ (covered by the blue dot), $\alpha=2$ (blue) and $\alpha=5$ (black). 
Lower dots: $\alpha=20$ (cyan) and $\alpha=175$ (purple, covering the $\alpha=0$ dot and partly also the cyan one)
\\Parameters: $\omega_0=4.0; a=m=1.0; \gamma=7.0; V=25$; $v_0 \in (-250,250)$.
}
\label{fig4}
\end{figure}
Here, a current emerges already for comparatively weak interactions ($\alpha=0.25$) 
and is accompanied, as before, by a significant increase
of the mean kinetic energy per particle compared to the noninteracting ($\alpha=0$) and the other nontransporting cases ($\alpha=20$; $\alpha=175$)
(Fig.~\ref{fig4}, inset b). 
The corresponding Fourier spectrum of the velocities of the ions (not shown) exhibits a continuous distribution. 
This provides the particle system with an enhanced flexibility to dynamically transfer energy between the individual particles. 
The emerging current is larger than that of the transporting regular excitation pattern discussed
($\langle \langle d\rangle \rangle \sim 0.2 \omega_0/k$) 
but still smaller than any single particle transport corresponding to a 1:1 resonance $\langle \langle d\rangle \rangle=\omega_0/(2k)$.
\paragraph*{Conclusions}
Opposite to the common assumption that directed currents represent a pure single particle effect that can survive 
interactions, we demonstrate here, that collective effects in repulsively 
power-law-interacting systems can induce directed particle currents even in parameter
regimes where corresponding noninteracting systems do not exhibit transport. 
These interaction induced currents could be detected
e.g. for cold thermal clouds of dipolar atoms and molecules \cite{lahaye09,lewenstein12} and interacting colloids \cite{bohlein12,bohlein12a} 
in optical lattices driven via standard techniques \cite{lewenstein12,windpassinger14} or alternatively 
with trapped ions additionally exposed to driven optical lattices \cite{pruttivarasin11}.
Our results might open a doorway towards the largly unexplored territory of interaction based transport phenomena in time dependent lattices. 
\bibliographystyle{apsrev}
\bibliography{./literature}

\begin{thebibliography}{43}
\expandafter\ifx\csname natexlab\endcsname\relax\def\natexlab#1{#1}\fi
\expandafter\ifx\csname bibnamefont\endcsname\relax
  \def\bibnamefont#1{#1}\fi
\expandafter\ifx\csname bibfnamefont\endcsname\relax
  \def\bibfnamefont#1{#1}\fi
\expandafter\ifx\csname citenamefont\endcsname\relax
  \def\citenamefont#1{#1}\fi
\expandafter\ifx\csname url\endcsname\relax
  \def\url#1{\texttt{#1}}\fi
\expandafter\ifx\csname urlprefix\endcsname\relax\def\urlprefix{URL }\fi
\providecommand{\bibinfo}[2]{#2}
\providecommand{\eprint}[2][]{\url{#2}}

\bibitem[{\citenamefont{Prost et~al.}(1994)\citenamefont{Prost, Chauwin,
  Peliti, and Ajdari}}]{prost94}
\bibinfo{author}{\bibfnamefont{J.}~\bibnamefont{Prost}},
  \bibinfo{author}{\bibfnamefont{J.-F.} \bibnamefont{Chauwin}},
  \bibinfo{author}{\bibfnamefont{L.}~\bibnamefont{Peliti}}, \bibnamefont{and}
  \bibinfo{author}{\bibfnamefont{A.}~\bibnamefont{Ajdari}},
  \bibinfo{journal}{Phys. Rev. Lett.} \textbf{\bibinfo{volume}{72}},
  \bibinfo{pages}{2652} (\bibinfo{year}{1994}).

\bibitem[{\citenamefont{Flach et~al.}(2000)\citenamefont{Flach, Yevtushenko,
  and Zolotaryuk}}]{flach00}
\bibinfo{author}{\bibfnamefont{S.}~\bibnamefont{Flach}},
  \bibinfo{author}{\bibfnamefont{O.}~\bibnamefont{Yevtushenko}},
  \bibnamefont{and}
  \bibinfo{author}{\bibfnamefont{Y.}~\bibnamefont{Zolotaryuk}},
  \bibinfo{journal}{Phys. Rev. Lett.} \textbf{\bibinfo{volume}{84}},
  \bibinfo{pages}{2358} (\bibinfo{year}{2000}).

\bibitem[{\citenamefont{Denisov et~al.}(2014)\citenamefont{Denisov, Flach, and
  H{\"a}nggi}}]{denisov14}
\bibinfo{author}{\bibfnamefont{S.}~\bibnamefont{Denisov}},
  \bibinfo{author}{\bibfnamefont{S.}~\bibnamefont{Flach}}, \bibnamefont{and}
  \bibinfo{author}{\bibfnamefont{P.}~\bibnamefont{H{\"a}nggi}},
  \bibinfo{journal}{Phys. Rep.} \textbf{\bibinfo{volume}{538}},
  \bibinfo{pages}{77} (\bibinfo{year}{2014}).

\bibitem[{\citenamefont{Magnasco}(1993)}]{magnasco93}
\bibinfo{author}{\bibfnamefont{M.}~\bibnamefont{Magnasco}},
  \bibinfo{journal}{Phys. Rev. Lett.} \textbf{\bibinfo{volume}{71}},
  \bibinfo{pages}{1477} (\bibinfo{year}{1993}).

\bibitem[{\citenamefont{Schanz et~al.}(2001)\citenamefont{Schanz, Otto,
  Ketzmerick, and Dittrich}}]{schanz01}
\bibinfo{author}{\bibfnamefont{H.}~\bibnamefont{Schanz}},
  \bibinfo{author}{\bibfnamefont{M.-F.} \bibnamefont{Otto}},
  \bibinfo{author}{\bibfnamefont{R.}~\bibnamefont{Ketzmerick}},
  \bibnamefont{and} \bibinfo{author}{\bibfnamefont{T.}~\bibnamefont{Dittrich}},
  \bibinfo{journal}{Phys. Rev. Lett.} \textbf{\bibinfo{volume}{87}},
  \bibinfo{pages}{070601} (\bibinfo{year}{2001}).

\bibitem[{\citenamefont{Reimann}(2002)}]{reimann02}
\bibinfo{author}{\bibfnamefont{P.}~\bibnamefont{Reimann}},
  \bibinfo{journal}{Phys. Rep.} \textbf{\bibinfo{volume}{361}},
  \bibinfo{pages}{57} (\bibinfo{year}{2002}).

\bibitem[{\citenamefont{H{\"a}nggi and Marchesoni}(2009)}]{hanggi09}
\bibinfo{author}{\bibfnamefont{P.}~\bibnamefont{H{\"a}nggi}} \bibnamefont{and}
  \bibinfo{author}{\bibfnamefont{F.}~\bibnamefont{Marchesoni}},
  \bibinfo{journal}{Rev. Mod. Phys.} \textbf{\bibinfo{volume}{81}},
  \bibinfo{pages}{387} (\bibinfo{year}{2009}).

\bibitem[{\citenamefont{Mennerat-Robilliard
  et~al.}(1999)\citenamefont{Mennerat-Robilliard, Lucas, Guibal, Tabosa,
  Jurczak, Courtois, and Grynberg}}]{mennerat99}
\bibinfo{author}{\bibfnamefont{C.}~\bibnamefont{Mennerat-Robilliard}},
  \bibinfo{author}{\bibfnamefont{D.}~\bibnamefont{Lucas}},
  \bibinfo{author}{\bibfnamefont{S.}~\bibnamefont{Guibal}},
  \bibinfo{author}{\bibfnamefont{J.}~\bibnamefont{Tabosa}},
  \bibinfo{author}{\bibfnamefont{C.}~\bibnamefont{Jurczak}},
  \bibinfo{author}{\bibfnamefont{J.-Y.} \bibnamefont{Courtois}},
  \bibnamefont{and} \bibinfo{author}{\bibfnamefont{G.}~\bibnamefont{Grynberg}},
  \bibinfo{journal}{Phys. Rev. Lett.} \textbf{\bibinfo{volume}{82}},
  \bibinfo{pages}{851} (\bibinfo{year}{1999}).

\bibitem[{\citenamefont{Schiavoni et~al.}(2003)\citenamefont{Schiavoni,
  Sanchez-Palencia, Renzoni, and Grynberg}}]{schiavoni03}
\bibinfo{author}{\bibfnamefont{M.}~\bibnamefont{Schiavoni}},
  \bibinfo{author}{\bibfnamefont{L.}~\bibnamefont{Sanchez-Palencia}},
  \bibinfo{author}{\bibfnamefont{F.}~\bibnamefont{Renzoni}}, \bibnamefont{and}
  \bibinfo{author}{\bibfnamefont{G.}~\bibnamefont{Grynberg}},
  \bibinfo{journal}{Phys. Rev. Lett.} \textbf{\bibinfo{volume}{90}},
  \bibinfo{pages}{094101} (\bibinfo{year}{2003}).

\bibitem[{\citenamefont{Salger et~al.}(2009)\citenamefont{Salger, Kling,
  Hecking, Geckeler, Morales-Molina, and Weitz}}]{salger09}
\bibinfo{author}{\bibfnamefont{T.}~\bibnamefont{Salger}},
  \bibinfo{author}{\bibfnamefont{S.}~\bibnamefont{Kling}},
  \bibinfo{author}{\bibfnamefont{T.}~\bibnamefont{Hecking}},
  \bibinfo{author}{\bibfnamefont{C.}~\bibnamefont{Geckeler}},
  \bibinfo{author}{\bibfnamefont{L.}~\bibnamefont{Morales-Molina}},
  \bibnamefont{and} \bibinfo{author}{\bibfnamefont{M.}~\bibnamefont{Weitz}},
  \bibinfo{journal}{Science} \textbf{\bibinfo{volume}{326}},
  \bibinfo{pages}{1241} (\bibinfo{year}{2009}).

\bibitem[{\citenamefont{Renzoni}(2009)}]{renzoni09}
\bibinfo{author}{\bibfnamefont{F.}~\bibnamefont{Renzoni}},
  \bibinfo{journal}{Adv. At. Mol. Opt. Phys.} \textbf{\bibinfo{volume}{57}},
  \bibinfo{pages}{1} (\bibinfo{year}{2009}).

\bibitem[{\citenamefont{Faucheux et~al.}(1995)\citenamefont{Faucheux, Bourdieu,
  Kaplan, and Libchaber}}]{faucheux95}
\bibinfo{author}{\bibfnamefont{L.~P.} \bibnamefont{Faucheux}},
  \bibinfo{author}{\bibfnamefont{L.~S.} \bibnamefont{Bourdieu}},
  \bibinfo{author}{\bibfnamefont{P.~D.} \bibnamefont{Kaplan}},
  \bibnamefont{and} \bibinfo{author}{\bibfnamefont{A.~J.}
  \bibnamefont{Libchaber}}, \bibinfo{journal}{Phys. Rev. Lett.}
  \textbf{\bibinfo{volume}{74}}, \bibinfo{pages}{1504} (\bibinfo{year}{1995}).

\bibitem[{\citenamefont{Marquet et~al.}(2002)\citenamefont{Marquet, Buguin,
  Talini, and Silberzan}}]{marquet02}
\bibinfo{author}{\bibfnamefont{C.}~\bibnamefont{Marquet}},
  \bibinfo{author}{\bibfnamefont{A.}~\bibnamefont{Buguin}},
  \bibinfo{author}{\bibfnamefont{L.}~\bibnamefont{Talini}}, \bibnamefont{and}
  \bibinfo{author}{\bibfnamefont{P.}~\bibnamefont{Silberzan}},
  \bibinfo{journal}{Phys. Rev. Lett.} \textbf{\bibinfo{volume}{88}},
  \bibinfo{pages}{168301} (\bibinfo{year}{2002}).

\bibitem[{\citenamefont{Matthias and M{\"u}ller}(2003)}]{matthias03}
\bibinfo{author}{\bibfnamefont{S.}~\bibnamefont{Matthias}} \bibnamefont{and}
  \bibinfo{author}{\bibfnamefont{F.}~\bibnamefont{M{\"u}ller}},
  \bibinfo{journal}{Nature} \textbf{\bibinfo{volume}{424}}, \bibinfo{pages}{53}
  (\bibinfo{year}{2003}).

\bibitem[{\citenamefont{Roeling et~al.}(2010)\citenamefont{Roeling, Germs,
  Smalbrugge, Geluk, de~Vries, Janssen, and Kemerink}}]{roeling10}
\bibinfo{author}{\bibfnamefont{E.~M.} \bibnamefont{Roeling}},
  \bibinfo{author}{\bibfnamefont{W.~C.} \bibnamefont{Germs}},
  \bibinfo{author}{\bibfnamefont{B.}~\bibnamefont{Smalbrugge}},
  \bibinfo{author}{\bibfnamefont{E.~J.} \bibnamefont{Geluk}},
  \bibinfo{author}{\bibfnamefont{T.}~\bibnamefont{de~Vries}},
  \bibinfo{author}{\bibfnamefont{R.~A.} \bibnamefont{Janssen}},
  \bibnamefont{and} \bibinfo{author}{\bibfnamefont{M.}~\bibnamefont{Kemerink}},
  \bibinfo{journal}{Nature Mater.} \textbf{\bibinfo{volume}{10}},
  \bibinfo{pages}{51} (\bibinfo{year}{2010}).

\bibitem[{\citenamefont{H{\"a}nggi}(2011)}]{hanggi11}
\bibinfo{author}{\bibfnamefont{P.}~\bibnamefont{H{\"a}nggi}},
  \bibinfo{journal}{Nature Mater.} \textbf{\bibinfo{volume}{10}},
  \bibinfo{pages}{6} (\bibinfo{year}{2011}).

\bibitem[{\citenamefont{Drexler et~al.}(2013)\citenamefont{Drexler, Tarasenko,
  Olbrich, Karch, Hirmer, M\"{u}ller, Gmitra, Fabian, Yakimova, Lara-Avila
  et~al.}}]{drexler13}
\bibinfo{author}{\bibfnamefont{C.}~\bibnamefont{Drexler}},
  \bibinfo{author}{\bibfnamefont{S.~a.} \bibnamefont{Tarasenko}},
  \bibinfo{author}{\bibfnamefont{P.}~\bibnamefont{Olbrich}},
  \bibinfo{author}{\bibfnamefont{J.}~\bibnamefont{Karch}},
  \bibinfo{author}{\bibfnamefont{M.}~\bibnamefont{Hirmer}},
  \bibinfo{author}{\bibfnamefont{F.}~\bibnamefont{M\"{u}ller}},
  \bibinfo{author}{\bibfnamefont{M.}~\bibnamefont{Gmitra}},
  \bibinfo{author}{\bibfnamefont{J.}~\bibnamefont{Fabian}},
  \bibinfo{author}{\bibfnamefont{R.}~\bibnamefont{Yakimova}},
  \bibinfo{author}{\bibfnamefont{S.}~\bibnamefont{Lara-Avila}},
  \bibnamefont{et~al.}, \bibinfo{journal}{Nature Nanotech.}
  \textbf{\bibinfo{volume}{8}}, \bibinfo{pages}{104} (\bibinfo{year}{2013}).

\bibitem[{\citenamefont{Duke and Austin}(1998)}]{duke98}
\bibinfo{author}{\bibfnamefont{T.~A.~J.} \bibnamefont{Duke}} \bibnamefont{and}
  \bibinfo{author}{\bibfnamefont{R.~H.} \bibnamefont{Austin}},
  \bibinfo{journal}{Phys. Rev. Lett.} \textbf{\bibinfo{volume}{80}},
  \bibinfo{pages}{1552} (\bibinfo{year}{1998}).

\bibitem[{\citenamefont{Chou et~al.}(1999)\citenamefont{Chou, Bakajin, Turner,
  Duke, Chan, Cox, Craighead, and Austin}}]{chou99}
\bibinfo{author}{\bibfnamefont{C.-F.} \bibnamefont{Chou}},
  \bibinfo{author}{\bibfnamefont{O.}~\bibnamefont{Bakajin}},
  \bibinfo{author}{\bibfnamefont{S.~W.} \bibnamefont{Turner}},
  \bibinfo{author}{\bibfnamefont{T.~A.} \bibnamefont{Duke}},
  \bibinfo{author}{\bibfnamefont{S.~S.} \bibnamefont{Chan}},
  \bibinfo{author}{\bibfnamefont{E.~C.} \bibnamefont{Cox}},
  \bibinfo{author}{\bibfnamefont{H.~G.} \bibnamefont{Craighead}},
  \bibnamefont{and} \bibinfo{author}{\bibfnamefont{R.~H.}
  \bibnamefont{Austin}}, \bibinfo{journal}{Proc. Natl. Acad. Sci.}
  \textbf{\bibinfo{volume}{96}}, \bibinfo{pages}{13762} (\bibinfo{year}{1999}).

\bibitem[{\citenamefont{Huang et~al.}(2003)\citenamefont{Huang, Cox, Austin,
  and Sturm}}]{huang03}
\bibinfo{author}{\bibfnamefont{L.~R.} \bibnamefont{Huang}},
  \bibinfo{author}{\bibfnamefont{E.~C.} \bibnamefont{Cox}},
  \bibinfo{author}{\bibfnamefont{R.~H.} \bibnamefont{Austin}},
  \bibnamefont{and} \bibinfo{author}{\bibfnamefont{J.~C.} \bibnamefont{Sturm}},
  \bibinfo{journal}{Anal. Chem.} \textbf{\bibinfo{volume}{75}},
  \bibinfo{pages}{6963} (\bibinfo{year}{2003}).

\bibitem[{\citenamefont{Mahmud et~al.}(2009)\citenamefont{Mahmud, Campbell,
  Bishop, Komarova, Chaga, Soh, Huda, Kandere-Grzybowska, and
  Grzybowski}}]{mahmud09}
\bibinfo{author}{\bibfnamefont{G.}~\bibnamefont{Mahmud}},
  \bibinfo{author}{\bibfnamefont{C.~J.} \bibnamefont{Campbell}},
  \bibinfo{author}{\bibfnamefont{K.~J.} \bibnamefont{Bishop}},
  \bibinfo{author}{\bibfnamefont{Y.~A.} \bibnamefont{Komarova}},
  \bibinfo{author}{\bibfnamefont{O.}~\bibnamefont{Chaga}},
  \bibinfo{author}{\bibfnamefont{S.}~\bibnamefont{Soh}},
  \bibinfo{author}{\bibfnamefont{S.}~\bibnamefont{Huda}},
  \bibinfo{author}{\bibfnamefont{K.}~\bibnamefont{Kandere-Grzybowska}},
  \bibnamefont{and} \bibinfo{author}{\bibfnamefont{B.~A.}
  \bibnamefont{Grzybowski}}, \bibinfo{journal}{Nature Phys.}
  \textbf{\bibinfo{volume}{5}}, \bibinfo{pages}{606} (\bibinfo{year}{2009}).

\bibitem[{\citenamefont{Di~Leonardo et~al.}(2010)\citenamefont{Di~Leonardo,
  Angelani, Dell’Arciprete, Ruocco, Iebba, Schippa, Conte, Mecarini,
  De~Angelis, and Di~Fabrizio}}]{dileonardo10}
\bibinfo{author}{\bibfnamefont{R.}~\bibnamefont{Di~Leonardo}},
  \bibinfo{author}{\bibfnamefont{L.}~\bibnamefont{Angelani}},
  \bibinfo{author}{\bibfnamefont{D.}~\bibnamefont{Dell’Arciprete}},
  \bibinfo{author}{\bibfnamefont{G.}~\bibnamefont{Ruocco}},
  \bibinfo{author}{\bibfnamefont{V.}~\bibnamefont{Iebba}},
  \bibinfo{author}{\bibfnamefont{S.}~\bibnamefont{Schippa}},
  \bibinfo{author}{\bibfnamefont{M.~P.} \bibnamefont{Conte}},
  \bibinfo{author}{\bibfnamefont{F.}~\bibnamefont{Mecarini}},
  \bibinfo{author}{\bibfnamefont{F.}~\bibnamefont{De~Angelis}},
  \bibnamefont{and}
  \bibinfo{author}{\bibfnamefont{E.}~\bibnamefont{Di~Fabrizio}},
  \bibinfo{journal}{Proc. Natl. Acad. Sci.} \textbf{\bibinfo{volume}{107}},
  \bibinfo{pages}{9541} (\bibinfo{year}{2010}).

\bibitem[{\citenamefont{Franken et~al.}(2012)\citenamefont{Franken, Swagten,
  and Koopmans}}]{franken12}
\bibinfo{author}{\bibfnamefont{J.~H.} \bibnamefont{Franken}},
  \bibinfo{author}{\bibfnamefont{H.~J.~M.} \bibnamefont{Swagten}},
  \bibnamefont{and} \bibinfo{author}{\bibfnamefont{B.}~\bibnamefont{Koopmans}},
  \bibinfo{journal}{Nature Nanotech.} \textbf{\bibinfo{volume}{7}},
  \bibinfo{pages}{499} (\bibinfo{year}{2012}).

\bibitem[{\citenamefont{Pethick and Smith}(2008)}]{pethick08}
\bibinfo{author}{\bibfnamefont{C.}~\bibnamefont{Pethick}} \bibnamefont{and}
  \bibinfo{author}{\bibfnamefont{H.}~\bibnamefont{Smith}},
  \emph{\bibinfo{title}{Bose-Einstein Condensation in Dilute Gases}}
  (\bibinfo{publisher}{Cambridge University Press, Cambridge},
  \bibinfo{year}{2008}).

\bibitem[{\citenamefont{Bloch et~al.}(2008)\citenamefont{Bloch, Dalibard, and
  Zwerger}}]{bloch08}
\bibinfo{author}{\bibfnamefont{I.}~\bibnamefont{Bloch}},
  \bibinfo{author}{\bibfnamefont{J.}~\bibnamefont{Dalibard}}, \bibnamefont{and}
  \bibinfo{author}{\bibfnamefont{W.}~\bibnamefont{Zwerger}},
  \bibinfo{journal}{Rev. Mod. Phys.} \textbf{\bibinfo{volume}{80}},
  \bibinfo{pages}{885} (\bibinfo{year}{2008}).

\bibitem[{\citenamefont{Gommers et~al.}(2005)\citenamefont{Gommers, Bergamini,
  and Renzoni}}]{gommers05}
\bibinfo{author}{\bibfnamefont{R.}~\bibnamefont{Gommers}},
  \bibinfo{author}{\bibfnamefont{S.}~\bibnamefont{Bergamini}},
  \bibnamefont{and} \bibinfo{author}{\bibfnamefont{F.}~\bibnamefont{Renzoni}},
  \bibinfo{journal}{Phys. Rev. Lett.} \textbf{\bibinfo{volume}{95}},
  \bibinfo{pages}{073003} (\bibinfo{year}{2005}).

\bibitem[{\citenamefont{Gommers et~al.}(2006)\citenamefont{Gommers, Denisov,
  and Renzoni}}]{gommers06}
\bibinfo{author}{\bibfnamefont{R.}~\bibnamefont{Gommers}},
  \bibinfo{author}{\bibfnamefont{S.}~\bibnamefont{Denisov}}, \bibnamefont{and}
  \bibinfo{author}{\bibfnamefont{F.}~\bibnamefont{Renzoni}},
  \bibinfo{journal}{Phys. Rev. Lett.} \textbf{\bibinfo{volume}{96}},
  \bibinfo{pages}{240604} (\bibinfo{year}{2006}).

\bibitem[{\citenamefont{Braun and Kivshar}(2004)}]{braun04}
\bibinfo{author}{\bibfnamefont{O.~M.} \bibnamefont{Braun}} \bibnamefont{and}
  \bibinfo{author}{\bibfnamefont{Y.~S.} \bibnamefont{Kivshar}},
  \emph{\bibinfo{title}{The Frenkel-Kontorova model: concepts, methods, and
  applications}} (\bibinfo{publisher}{Springer}, \bibinfo{year}{2004}).

\bibitem[{\citenamefont{Quintero}(2014)}]{cuevas14}
\bibinfo{author}{\bibfnamefont{N.~R.} \bibnamefont{Quintero}}
  (\bibinfo{publisher}{Springer International Publishing},
  \bibinfo{year}{2014}), vol.~\bibinfo{volume}{10} of
  \emph{\bibinfo{series}{Nonlinear Systems and Complexity}}, pp.
  \bibinfo{pages}{131--154}.

\bibitem[{\citenamefont{Lahaye et~al.}(2009)\citenamefont{Lahaye, Menotti,
  Santos, Lewenstein, and Pfau}}]{lahaye09}
\bibinfo{author}{\bibfnamefont{T.}~\bibnamefont{Lahaye}},
  \bibinfo{author}{\bibfnamefont{C.}~\bibnamefont{Menotti}},
  \bibinfo{author}{\bibfnamefont{L.}~\bibnamefont{Santos}},
  \bibinfo{author}{\bibfnamefont{M.}~\bibnamefont{Lewenstein}},
  \bibnamefont{and} \bibinfo{author}{\bibfnamefont{T.}~\bibnamefont{Pfau}},
  \bibinfo{journal}{Rep. Prog. Phys.} \textbf{\bibinfo{volume}{72}},
  \bibinfo{pages}{126401} (\bibinfo{year}{2009}).

\bibitem[{\citenamefont{Lewenstein et~al.}(2012)\citenamefont{Lewenstein,
  Sanpera, and Ahufinger}}]{lewenstein12}
\bibinfo{author}{\bibfnamefont{M.}~\bibnamefont{Lewenstein}},
  \bibinfo{author}{\bibfnamefont{A.}~\bibnamefont{Sanpera}}, \bibnamefont{and}
  \bibinfo{author}{\bibfnamefont{V.}~\bibnamefont{Ahufinger}},
  \emph{\bibinfo{title}{Ultracold Atoms in Optical Lattices: Simulating quantum
  many-body systems}} (\bibinfo{publisher}{Oxford University Press},
  \bibinfo{year}{2012}).

\bibitem[{\citenamefont{Bohlein and Bechinger}(2012)}]{bohlein12a}
\bibinfo{author}{\bibfnamefont{T.}~\bibnamefont{Bohlein}} \bibnamefont{and}
  \bibinfo{author}{\bibfnamefont{C.}~\bibnamefont{Bechinger}},
  \bibinfo{journal}{Phys. Rev. Lett.} \textbf{\bibinfo{volume}{109}},
  \bibinfo{pages}{058301} (\bibinfo{year}{2012}).

\bibitem[{\citenamefont{Pruttivarasin et~al.}(2011)\citenamefont{Pruttivarasin,
  Ramm, Talukdar, Kreuter, and H{\"a}ffner}}]{pruttivarasin11}
\bibinfo{author}{\bibfnamefont{T.}~\bibnamefont{Pruttivarasin}},
  \bibinfo{author}{\bibfnamefont{M.}~\bibnamefont{Ramm}},
  \bibinfo{author}{\bibfnamefont{I.}~\bibnamefont{Talukdar}},
  \bibinfo{author}{\bibfnamefont{A.}~\bibnamefont{Kreuter}}, \bibnamefont{and}
  \bibinfo{author}{\bibfnamefont{H.}~\bibnamefont{H{\"a}ffner}},
  \bibinfo{journal}{New J. Phys.} \textbf{\bibinfo{volume}{13}},
  \bibinfo{pages}{075012} (\bibinfo{year}{2011}).

\bibitem[{\citenamefont{Liebchen et~al.}(2012)\citenamefont{Liebchen, Diakonos,
  and Schmelcher}}]{liebchen12}
\bibinfo{author}{\bibfnamefont{B.}~\bibnamefont{Liebchen}},
  \bibinfo{author}{\bibfnamefont{F.~K.} \bibnamefont{Diakonos}},
  \bibnamefont{and}
  \bibinfo{author}{\bibfnamefont{P.}~\bibnamefont{Schmelcher}},
  \bibinfo{journal}{New J. Phys.} \textbf{\bibinfo{volume}{14}},
  \bibinfo{pages}{103032} (\bibinfo{year}{2012}).

\bibitem[{\citenamefont{Frenkel and Smit}(2002)}]{frenkel01}
\bibinfo{author}{\bibfnamefont{D.}~\bibnamefont{Frenkel}} \bibnamefont{and}
  \bibinfo{author}{\bibfnamefont{B.}~\bibnamefont{Smit}},
  \emph{\bibinfo{title}{Understanding molecular simulation: from algorithms to
  applications}} (\bibinfo{publisher}{Academic press, San Diego},
  \bibinfo{year}{2002}).

\bibitem[{\citenamefont{Dormand and Prince}(1980)}]{dormand80}
\bibinfo{author}{\bibfnamefont{J.}~\bibnamefont{Dormand}} \bibnamefont{and}
  \bibinfo{author}{\bibfnamefont{P.}~\bibnamefont{Prince}},
  \bibinfo{journal}{J. Comput. Appl. Math.} \textbf{\bibinfo{volume}{6}},
  \bibinfo{pages}{19 } (\bibinfo{year}{1980}).

\bibitem[{\citenamefont{Callen and Welton}(1951)}]{callen51}
\bibinfo{author}{\bibfnamefont{H.~B.} \bibnamefont{Callen}} \bibnamefont{and}
  \bibinfo{author}{\bibfnamefont{T.~A.} \bibnamefont{Welton}},
  \bibinfo{journal}{Phys. Rev.} \textbf{\bibinfo{volume}{83}},
  \bibinfo{pages}{34} (\bibinfo{year}{1951}).

\bibitem[{\citenamefont{Zwanzig}(2001)}]{zwanzig01}
\bibinfo{author}{\bibfnamefont{R.}~\bibnamefont{Zwanzig}},
  \emph{\bibinfo{title}{Nonequilibrium statistical mechanics}}
  (\bibinfo{publisher}{Oxford University Press}, \bibinfo{year}{2001}).

\bibitem[{\citenamefont{Barbi and Salerno}(2000)}]{salerno00}
\bibinfo{author}{\bibfnamefont{M.}~\bibnamefont{Barbi}} \bibnamefont{and}
  \bibinfo{author}{\bibfnamefont{M.}~\bibnamefont{Salerno}},
  \bibinfo{journal}{Phys. Rev. E} \textbf{\bibinfo{volume}{62}},
  \bibinfo{pages}{1988} (\bibinfo{year}{2000}).

\bibitem[{\citenamefont{Pikovsky et~al.}(2001)\citenamefont{Pikovsky,
  Rosenblum, and Kurths}}]{pikovsky01}
\bibinfo{author}{\bibfnamefont{A.}~\bibnamefont{Pikovsky}},
  \bibinfo{author}{\bibfnamefont{M.}~\bibnamefont{Rosenblum}},
  \bibnamefont{and} \bibinfo{author}{\bibfnamefont{J.}~\bibnamefont{Kurths}},
  \emph{\bibinfo{title}{A universal concept in nonlinear sciences}}
  (\bibinfo{publisher}{Cambridge University Press}, \bibinfo{year}{2001}).

\bibitem[{\citenamefont{Baker and Gollub}(1996)}]{baker96}
\bibinfo{author}{\bibfnamefont{G.~L.} \bibnamefont{Baker}} \bibnamefont{and}
  \bibinfo{author}{\bibfnamefont{J.~P.} \bibnamefont{Gollub}},
  \emph{\bibinfo{title}{Chaotic dynamics: an introduction}}
  (\bibinfo{publisher}{Cambridge University Press}, \bibinfo{year}{1996}).

\bibitem[{\citenamefont{Bohlein et~al.}(2012)\citenamefont{Bohlein, Mikhael,
  and Bechinger}}]{bohlein12}
\bibinfo{author}{\bibfnamefont{T.}~\bibnamefont{Bohlein}},
  \bibinfo{author}{\bibfnamefont{J.}~\bibnamefont{Mikhael}}, \bibnamefont{and}
  \bibinfo{author}{\bibfnamefont{C.}~\bibnamefont{Bechinger}},
  \bibinfo{journal}{Nature Mater.} \textbf{\bibinfo{volume}{11}},
  \bibinfo{pages}{126} (\bibinfo{year}{2012}).

\bibitem[{\citenamefont{Windpassinger and Sengstock}(2013)}]{windpassinger14}
\bibinfo{author}{\bibfnamefont{P.}~\bibnamefont{Windpassinger}}
  \bibnamefont{and}
  \bibinfo{author}{\bibfnamefont{K.}~\bibnamefont{Sengstock}},
  \bibinfo{journal}{Rep. Prog. Phys.} \textbf{\bibinfo{volume}{76}},
  \bibinfo{pages}{086401} (\bibinfo{year}{2013}).

\end{thebibliography}

\end{document}